\documentclass[twocolumn,twoside,slac]{revtex4}
\usepackage{graphicx}
\usepackage{fancyhdr}
\begin{document}

\title{AliEnFS - a Linux File System for the AliEn Grid Services}

%

\author{Andreas-J. Peters }
\affiliation{CERN, CH 1211 Geneva 23, Switzerland}
\author{P. Saiz}
\affiliation{CERN, CH 1211 Geneva 23, Switzerland}
\affiliation{University of the West England, Frenchay Campus
  Coldharbour Lane, Bristol BS161QY, U.K.}
\author{P. Buncic}
\affiliation{Institut f\"ur Kernphysik, Universit\"at Frankfurt,
  August-Euler-Strasse 6, 60486 Frankfurt am Main, Germany}
\affiliation{CERN, CH 1211 Geneva 23, Switzerland}

\begin{abstract}
Among the services offered by the AliEn (ALICE Environment
http://alien.cern.ch) Grid framework there is a virtual file catalogue to
allow transparent access to distributed data-sets using various file
transfer protocols. $alienfs$ (AliEn File System) integrates the AliEn file
catalogue as a new file system type into the Linux kernel using LUFS, a
hybrid user space file system framework (Open Source
http://lufs.sourceforge.net). LUFS uses a special kernel interface level
called VFS (Virtual File System Switch) to communicate via a generalised
file system interface to the AliEn file system daemon. The AliEn 
framework is used for authentication, catalogue browsing, file
registration and read/write transfer operations. A C++ API implements the
generic file system operations. The goal of AliEnFS is to allow
users easy interactive access to a worldwide distributed virtual file system using
familiar shell commands (f.e. cp,ls,rm ...)\\
The paper discusses general aspects of Grid File Systems, the AliEn
implementation and present and future developments for the AliEn Grid
File System.
\end{abstract}

\maketitle

\thispagestyle{fancy}


\section*{Physicist requirements for Grid Services}
The basic Grid services in new LHC era experiments \cite{LHC} are distributed
data storage and computing for high performance access and parallel processing of 
experimental data. Two fundaments have
to be provided in this environment:
\begin{itemize}
\item
  a Global Grid File System
\item
  a Global Queue System and/or a Global Analysis Framework
\end{itemize}

In the following sections we will discuss the layout of a Global Grid
File System in general and the concrete implementation in the AliEn Grid.

\section{The Global Grid File System}

The Global Grid File System has to glue multiple different types of
storage systems and data transport protocols into one virtual
file system. A common abstract user access interface has to be
provided while concrete implementations can be hidden.\\
It should support access privileges and ownership of directories and
files for individual users (user), group of users (group) or all members in the
community (others). Moreover meta data should be tagged to Grid Files for
indexing. In the context of a conventual file system, meta data plays a
minor role. \\
The file system representation needs:
\begin{itemize}
\item
  shell(-like) commands (ls, mkdir, cp, rm ...)
\item
  application file access
\end{itemize}
A unix-like command line interface are nowadays standard
requirements, which are important for interactive sessions.\\
On the other hand application file access plays a more crucial role in the context
of HENP applications and will be discussed in the following sections.

\subsection{Constituents of a Grid File System} 
We can distinguish three spheres for the implementation of a Grid File
System:
\begin{itemize}
\item
  Hierarchical Virtual Directory Structure
\item
  Data Storage Elements
\item
  Data Transfer Layer
\end{itemize}

\subsubsection {Hierarchical Virtual Directory Structure}
The user interface of distributed file storages is based on a virtual directory
structure, which connects virtual (in the following referred to as logical) file
names to physical locations and their access methods. These LFNs
(Logical File Names) connect to PFNs (Physical File Names) in a common
URL format:

\begin{eqnarray}
 <\mathrm{LFN}> &\longrightarrow& <\mathrm{PFN}> \nonumber \\
 <\mathrm{PFN}> &=& <\mathrm{protocol}>:// \nonumber\\
                & &
 <\mathrm{host}>:<\mathrm{port}>\/<\mathrm{direntry}> \nonumber
\end{eqnarray}

Logical File Names are constructed in a file system-like notation: $
\mathrm{grid://}[dir]/[subdir]/../[filename]$. Directories and subdirectories are
connected like inodes in a conventional file system.\\

\subsubsection {Data Storage Elements}
Data Storage Elements (SE) implement the functionality of persistent data
storage. A SE provides the namespace and storage access for PFNs. At
the backend SEs use conventional
Disk File Systems or Mass Storage Systems like Castor \cite{Castor},
DCache \cite{DCache},
HPSS \cite{HPSS} etc. \\
Physical locations and names of files are given by the
configuration of a SE and have no connection to LFN names in
the virtual directory structure.

\subsubsection {Data Transfer Layer}
\begin{figure}
\includegraphics[width=65mm]{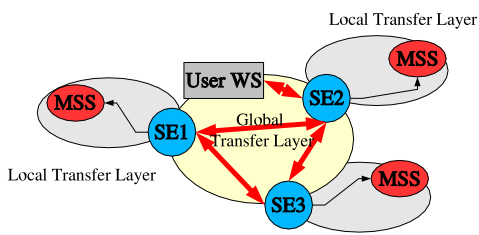}
\caption{Global and local transfer layer in Grid File Systems.}
\label{transferlayer}
\end{figure}

The Data Transfer Layer can be divided into a local and a global
component (see figure \ref{transferlayer}). The local data transfer layer describes the data access of
SEs to the local used storage system via local access
protocols like POSIX I/O \cite{POSIX}, ROOT I/O \cite{ROOT}, RFIO
\cite{Castor}, DCache \cite{DCache} etc. The global data transfer
component specifies data transfer between different SEs or between
SEs and remote users/applications. Moreover one can distinguish
scheduled and interactive access, while in the context of a file system we
will consider mostly interactive (immediate) file access

\subsection{Implementation Approaches for Application File Access}
Mainly three different approaches for a Grid file system implementation
exist:
\begin{itemize}
\item
  a Virtual File System as a kernel module in the operation system
\item
  a Bypass library which overloads standard POSIX I/O commands
\item
  a Grid application interface library which has to be linked into
  Grid connected applications
\end{itemize}
\subsubsection{VFS modules}
The implementation of a Virtual File System module, as it is common in
Operating Systems like LINUX, is the preferred solution for
interactive user access 
 \cite{VFS}. The user can access the Grid File System in a familiar
way like every standard file system (EXT3/{\bf afs} etc.) with shell
commands, POSIX I/O etc.\\
A difficulty with the implementation of such modules is the
implementation of Grid connectivity in a kernel environment. A
solution to this problem will be discussed under \ref{LUFSsection}. 

\subsubsection{Bypass and Application Interface Libraries (API)}
Since
a kernel module needs a bigger installation effort and requires
high stability and availability, the preferred method for production
farm computers is a Bypass or an API library. While a Bypass library
may dedicate a computer completely to one Grid implementation, a
connection of Grid functionality to programs by use of an API library
allows the most flexible and stable use of Grid File System functionality. 

\section{Implementation of the Global Grid File System in AliEn}
AliEn (ALICE {\it Environment}) is an implementation of distributed
computing infrastructure (Grid framework) designed to simulate,
reconstruct and analyze data of the ALICE experiment. It is built on
top of Internet standards for information exchange and authentication
(SOAP \cite{SOAP},PKI) and common Open Source components. It provides a file catalogue and services for authentication, job execution, file
transport, performance monitoring and event logging. The AliEn core
services are implemented using perl. Further reading can be done under
\cite{alienpaper} and \cite{predragpaper}.
\subsection{The virtual file catalogue in AliEn}
\begin{figure}
\includegraphics[width=65mm]{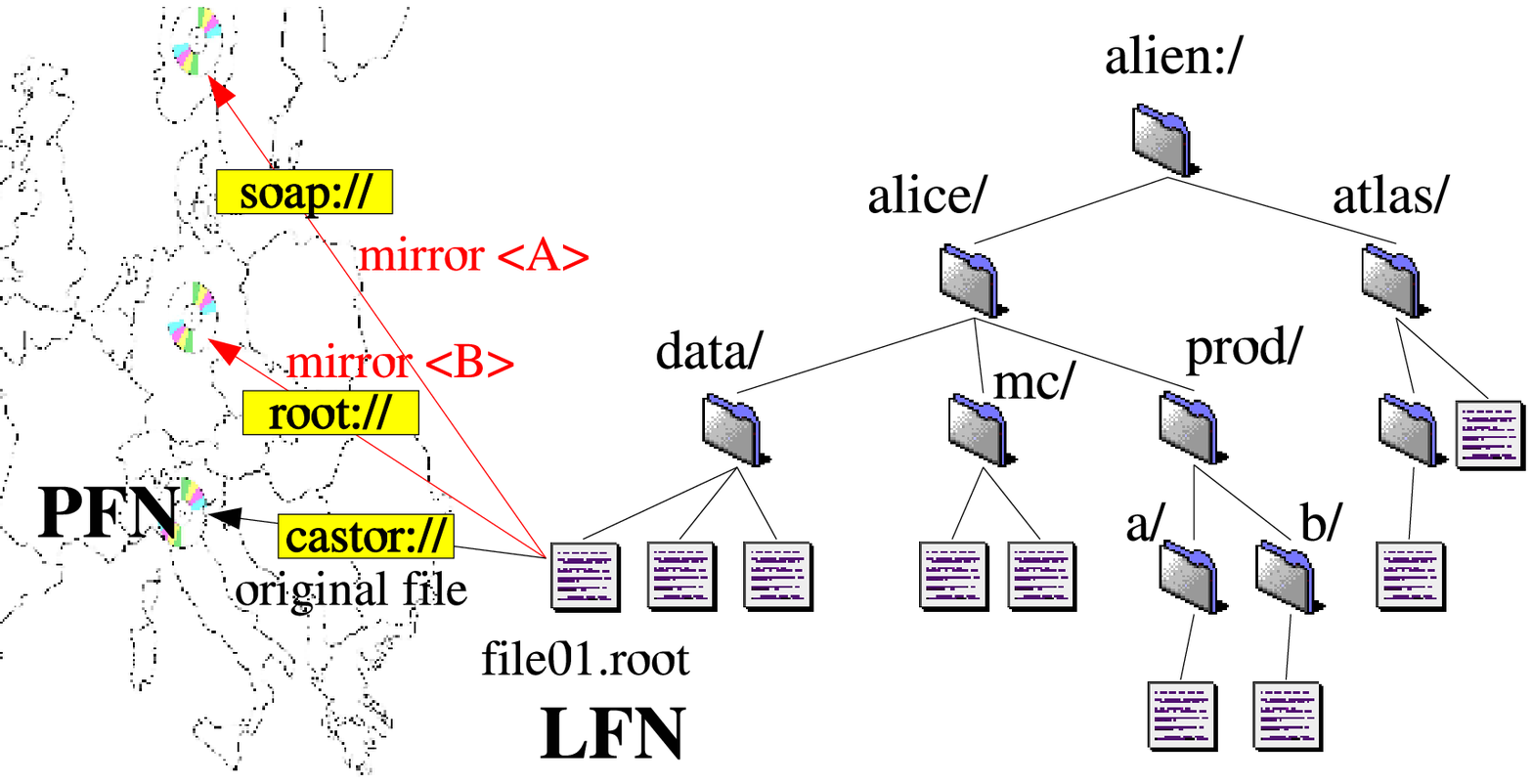}
\caption{Hierarchical structure of the virtual file catalogue in AliEn.}
\label{catalogue}
\end{figure}

The virtual directory structure for the AliEn Grid File System is
provided by the AliEn file catalogue (see figure \ref{catalogue}). It is implemented using
a database independent interface which connects at the moment to MySQL databases \cite{MYSQL}. In terms of database labeling, virtual directories are
represented by tables in the database. Subdirectories are connected to
directories through sub-table entries. For performance gain different directory branches can be
assigned to different database servers.\\
The base names of LFNs are
represented as entries in directory tables. These entries contain
beside the LFN itself the {\it master} PFN with information about the
physical location and the access protocol of the specific
files. Locations of replicated files are not stored in the LFN entries
itself but in a dedicated table containing LFNs and replica locations.\\
For fast response an additional database table contains all LFNs and
the corresponding sub-table references, where a specific LFN can be
found. Each LFN entry contains moreover size, owner, group and access
privilege information.\\
Additional information about the virtual file catalogue can be found under
\cite{alienpaper} and \cite{PABLOPAPER}.

\subsection{Data Storage Elements in AliEn}
\subsubsection{Basic Storage Element}
The Basic Storage Element implements the storage, access and PFN labeling functionality
for files. It deals only with complete files and does not support partial
file access. It provides a plug-in architecture to support any
kind of storage systems. Therefore standard directory and file
manipulation functions are implemented in plug-in modules (see table \ref{SEplugins}).
\begin{table}[t]
\begin{center}
\caption{Plug-in Functions of the Basic Storage Element.}
\begin{tabular}{|l|}
\hline \textbf{mkdir} \\
 creation of new directory in the storage system \\
\hline \textbf{link} \\
 symbolic link of two files in the storage system \\
\hline \textbf{cp} \\
 copy of a file from/into the storage system \\
\hline \textbf{mv} \\
 rename of a file inside the storage system\\
\hline \textbf{rm} \\
 deletion of a file inside the storage system\\
\hline \textbf{sizeof} \\
 file size determination inside the storage system\\
\hline \textbf{url} \\
 URL file name in accordance with PFN definition\\
\hline \textbf{lslist} \\
 file name and size listing of all files inside \\the storage system \\
\hline
\end{tabular}
\label{SEplugins}
\end{center}
\end{table}


The SE configuration defines the plug-in module to use. The currently
supported storage systems can be seen in table \ref{SEmodules}.
\begin{table}[t]
\begin{center}
\caption{Mass Storage Systems supported by AliEn.}
\begin{tabular}{|l|}
\hline \textbf{ADSM} \cite{ADSM}\\ ADSTAR Distributed Storage Manager system\\
\hline \textbf{AliEn} \cite{AliEn}\\ a virtual AliEn SE for
$\mathrm{LFN}\rightarrow \mathrm{LFN}$ symb. links\\
\hline \textbf{Castor} \cite{Castor}\\ CERN Mass Storage System\\
\hline \textbf{DB} \\ Database Storage System \\
\hline \textbf{DMF} \cite{DMF}\\ Data Migration Facility system \\
\hline \textbf{EDG} \cite{EDG}\\ European Data Grid SE interface\\
\hline \textbf{File} \\ SE for file access in locally mounted filesystems\\
\hline \textbf{HPSS} \cite{HPSS}\\  High Performance Storage System using RFIO access\\
\hline \textbf{HSI} \cite{HSI}\\ High Performance Storage System unsing HSI access\\
\hline \textbf{HTTP} \\ web server storage element\\
\hline \textbf{SOAP} \cite{SOAP}\\ SOAP server storage element \\
\hline
\end{tabular}
\label{SEmodules}
\end{center}
\end{table}


Files are stored in a local disk cache for performance
improvement in case of repetitive file access.

\subsubsection{The Logical Volume Manager}
The Logical Volume Manager allows to merge several storage system volumes of
the same kind into one logical storage volume. This is especially
useful to cluster a set of raid disk systems or hard disks into one logical
volume. For each volume a mount point, disk space and a file lifetime can be
defined. The Logical Volume Manager keeps track of used and available
disk space and blocks additional storage save operations, if the SE size
is exceeded. Expired files can be erased. Using the {\it lslist} function of the SE plug-in modules, it
allows (re-)synchronisation of existing data in the storage
volumes.
\subsubsection{I/O Server}
To extend the functionality of the Basic Storage Element, which
supports only access of complete files, I/O daemons can be started permanently or on
demand to support partial file access. The possible protocols are
described in the following subsection.
\begin{figure}
\includegraphics[width=65mm]{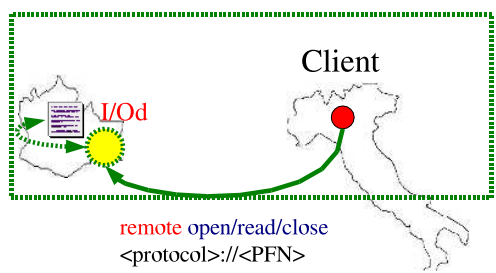}
\caption{File Access through remote I/O daemons.}
\label{RIO}
\end{figure}

\subsection{Data Transfer Layer in AliEn}
\label{protocols}
AliEn uses at the moment three mechanism for file transfer. Small files
can be accessed using the SOAP \cite{SOAP} protocol, if supported by the
SE. Otherwise a Transfer Broker exists to move or replicate files, which
have been entered into the Transfer Queue. The Transfer is
executed by File Transfer Daemons, which reside on SEs using
a bbFTP \cite{BBFTP} server and client. Since the transfers are scheduled from a
queue, this has to be considered as an asynchronous mechanism.\\ For interactive
and partial file access, AliEn uses I/O servers, which can be started
on demand or permanently at SEs. Possible servers could be
f.e. gridFTP \cite{GRIDFTP},
rootd \cite{ROOT}, sshd \cite{SSH}, rfiod \cite{Castor}, a self-developed server etc. The best
choice is still in the evaluation phase (see section \ref{future}).

\subsubsection{Access pattern and strategies for partial file access}
The AliEn SE supports two operations for file access:
\begin{itemize}
\item
  $read$ (offset,size)
\item
  $write~once$ (streamed)
\end{itemize}
While the $read$ operation allows to read files at arbitrary
positions, the $write$ operation allows only streaming without offset
change between concurrent $write$ operations. An $open$ for  $write$
can be done only on not existent files, what is often refered to as
{\it write once} modus. This guarantees the consistency of replicated or
cached data. \\

For the $read$ operation two access strategies are supported:
\begin{itemize}
\item
  partial file transfer via remote I/O server according to $read$
  requests
\item
  local download of the complete file and partial file access from the
  local disk
\end{itemize}
Accordingly the strategies for $write$ operations are:
\begin{itemize}
\item
  partial file transfers to remote I/O server according to $write$
  requests
\item
  $write$ to local disk cache and complete file transfer after end of
  $write$ operations ($close$).
\end{itemize}
Evidently the application has to select the appropriate access
strategy.\\

\subsubsection{Operation sequence for file access}
\begin{figure*} [htbp]
\includegraphics[width=140mm]{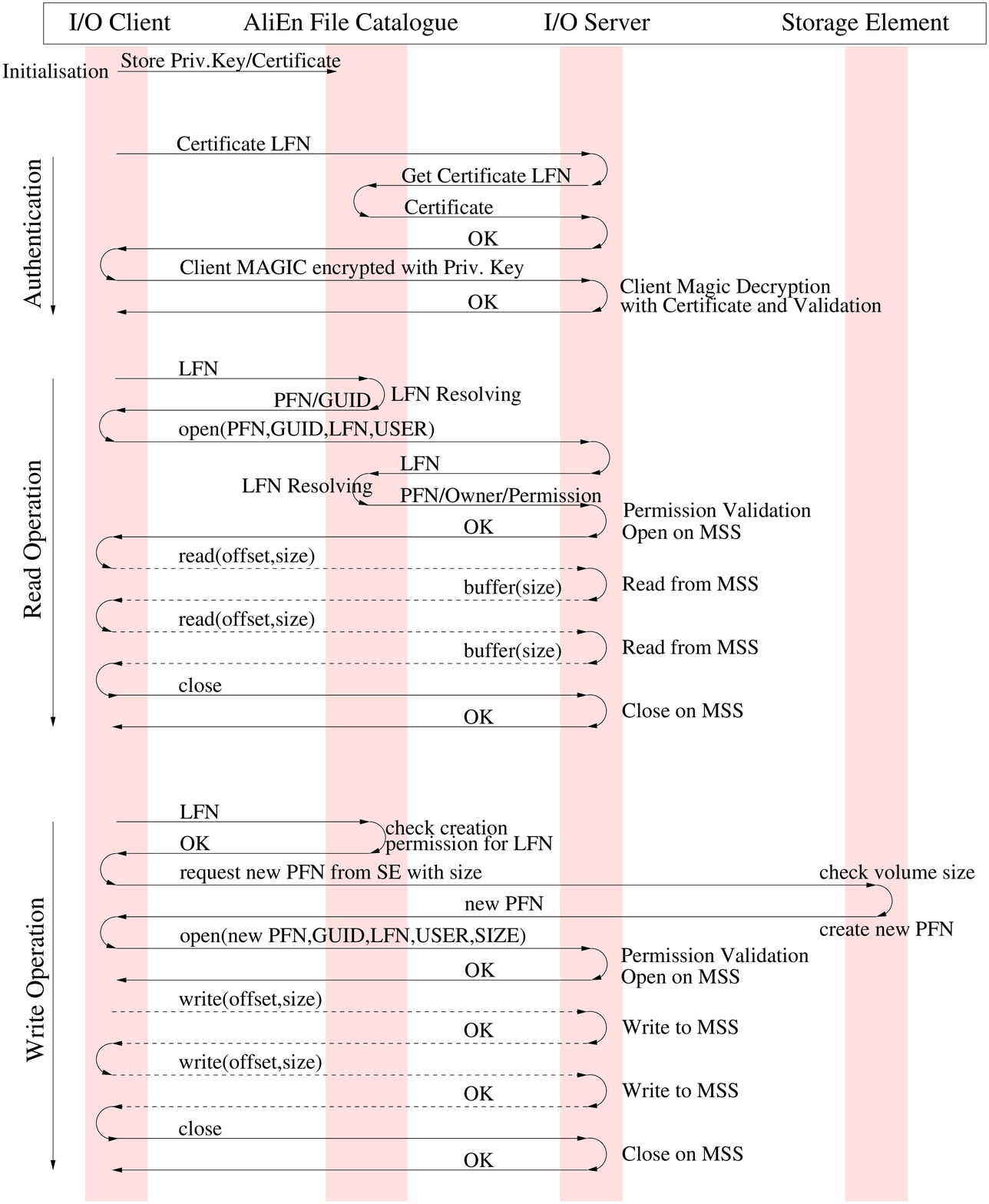}
\caption{Flow diagram for {\it Authentication},{\it Read} and {\it
    Write} file access.}
\label{IOflow}
\end{figure*}

The flow diagram for file access is shown in figure \ref{IOflow}.
The operation sequence for a $read$ operation starts with the LFN
resolving. A dedicated AliEn service resolves all locations of a file and selects the
{\it best} access location and protocol. For the time being this is done
with a trivial matching of the closest location. For the $open$
operation, a connection to the remote I/O server is established and
possibly (depending on the access strategy) downloaded as a complete file. 
With the $open$ command the user permissions are checked on client
and server side. The $close$ command closes a remote connection or
the local file.
The operation sequence for a $write$ operation starts on $open$
with a privilege check for the creation of a new LFN entry in the
given directory. If this is passed, the SE is contacted to provide a
new PFN. It also examines, if there is enough space to store the file.
At the end of the $open$ sequence, a connection is opened 
to the remote I/O server. With the $close$ operation, the file size
in the SE is validated and the LFN/PFN pair inserted in the file
catalogue. \\
The client has to authenticate in both cases to the I/O server using
the appropriate authentication method for the used protocol.

\subsection{The Grid API of AliEN}
The above described file access methods are implemented in the AliEn
C++ API Library. Moreover the API library contains all needed methods 
for authentication and exchange with Grid services.\\
Since all physical directory creation and deletion
operations are up to the SE, the API provides directory creation, browsing,
deletion and privilege/owner manipulation of files and directories
only for LFNs. Physical files can be created or accessed through so called {\it generic} I/O commands:
{\vspace{2mm}
{ \tt
  ~~~\hspace{5mm}FDTYPE~~~{genericopen}~('LFN');\\
  ~~~\hspace{5mm}ssize\_t~~{genericread}~(FDTYPE handle, void *buffer,loff\_t offset,size\_t size);\\
  ~~~\hspace{5mm}ssize\_t~~{genericwrite}(FDTYPE handle, void *buffer,loff\_t offset,size\_t size);\\
  ~~~\hspace{5mm}int~~~~~~{genericclose}(FDTYPE);\\
  ~~~\hspace{5mm}int~~~~~~{genericsync}~(FDTYPE);\\
}
\hspace{2mm}}\\
The API itself keeps track after an $open$ of the needed
remote connections or local file descriptors to proceed with the generic
$read$ or $write$ operations. These informations are stored in a
generic list of open files. Each $open$ assigns a file handle
(FDTYPE) to be
used in consecutive I/O operations to associate with entries in the
generic list.

\subsection{Virtual File System Implementation for AliEn under Linux}
\label{LUFSsection}
As already mentioned, the problematic of a virtual file system
implementation under Linux is based on the need of high level libraries
like the AliEn C++ API \cite{API}. In a VFS module in the kernel
environment only kernel functions are known and shared libraries
cannot be linked.\\
These technical problem can be solved by splitting the VFS module into
two parts: one in kernel space and one in user space. The user space
part can be linked to any external library. All kernel requests have
to be redirected from the core VFS kernel module to the user space
module.\\

\subsubsection{LUFS - Linux Userland File System}
\begin{figure}
\includegraphics[width=65mm]{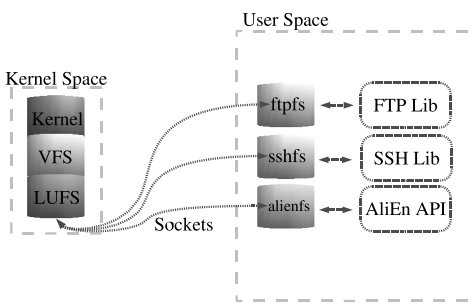}
\caption{LUFS communication between kernel module and user space daemons.}
\label{LUFS}
\end{figure}

The open source project LUFS ('Linux Userland File System') \cite{LUFSpage}
offers a modular file system plug-in architecture with a generic
LUFS VFS kernel module, which communicates with various user space
file system modules (see figure \ref{LUFS}).\\
The recent version supports Linux kernel V2.4 and V2.5 with large file support.\\
A couple of file system modules come with LUFS like $ftpfs$,
$sshpfs$, $cardfs$ et al. $ftpfs$ e.g. utilises the $ftp$ client protocol
to mount a file system from an $ftp$ server. In almost the same manner
the AliEn Grid file system module $alienfs$ is implemented. It uses
the AliEn API library for directory and file handling.\\
The $mount$ command has to be executed by each individual user. During
the $mount$ execution the Grid authentication is done only once for the user
space thread. In case of a broken connection, the
connection to the Grid is automatically rebuilt. Therefore it is
recommended to choose a key- or proxy-certificate-based authentication
method.\\
Some extensions to the $generic$ library calls have to be made, since
LUFS uses in $read$ and $write$ operations always the LFN as a file
descriptor and the offset position is always given as an argument in
the I/O call. \\
The implementation of the kernel module is done in such a way that it 
redirects VFS calls via UNIX domain sockets to a LUFS
daemon which runs in user space. This daemon loads according to the
given mount option a shared library module for the requested file
system module. The LUFS daemon provides also directory caching with
configurable lifetime for performance increase in repetitive $stat$
calls.\\
The user/group translation from the AliEn file catalogue uses the
local $passwd$ and $group$ files for
ownership translation.\\
In case of a write to the Grid file system, the assigned
default SE is used for file storage. The user can select an
alternative SE by appending ''@SE-Name'' to the destination file.\\
In the future file meta data can be made visible in the file
system as virtual {\it extra} files, which return in case of a read
operation the meta data tag list.\\
LUFS also supports $automount$, which is a desirable functionality for
a Grid File System.\\
\subsubsection{{\it gridfs} - a generalised Grid File System Plug-in Architecture}
\begin{figure}
\includegraphics[width=65mm]{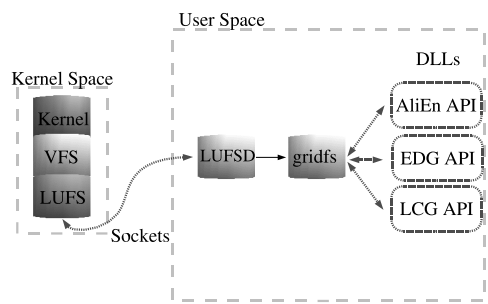}
\caption{LUFS communication between kernel module, user space daemons
  and {\it gridfs} plug-in library.}
\label{LUFS2}
\end{figure}

To provide a more flexible framework, a more general {\it gridfs} module
has been developed. This module allows dynamic loading of Grid API
libraries and enables to use the same
LUFS module for various Grid platforms (see figure \ref{LUFS2}). \\
To introduce a new Grid File System, one has to implement a template
class, which is the base for a shared library. By executing the mount
command, the library name of the plug-in class has to be passed as an
option to the LUFS {\it gridfs} module. Additional needed Grid API
libraries can also be given as a mount option and will be loaded
dynamically from the plug-in module. The AliEn plug-in is at present
in the evaluation and debugging phase.

\subsubsection{Provisos against Bypass libraries}
The Bypass library mechanism has not yet been tested since there are
some obvious objections. A Bypass library redirects system calls to
the Grid API library. This has some performance implications for
non-Grid applications.\\
Another difficult item is, how to avoid repeated authentication of
individual users with every system call.\\
Therefore the Bypass library is at present not considered to be an applicable
solution for the AliEn Grid File System.

\section{Present and future developments of the AliEn Grid File
  System}
\label{future}
The crucial point of a Grid File System resides not in the
  end points like the user access through the VFS or the SE itself
  but in the $global$ transport layer. Some effort has been spent in
  the recent time to establish a new efficient transport layer.\\
  Requirements for a flexible, efficient and safe transport
  architecture are:
\begin{itemize}
\item secure Grid User Authentication
\item {\bf A}ccess {\bf C}ontrol {\bf L}ists
\item optional data encryption
\item optimization for high- and low-bandwidth transfers
\item 'Network Weather Service':
  \begin{itemize}
    \item dynamic bandwidth regulation per connection
    \item transfer (re-)routing through cache servers and high-speed
    connections
  \end{itemize}
\item load balancing
  \begin{itemize}
  \item on-site distributed cache server 
  \item on-site distributed I/O server   
  \end{itemize}
\item safe operation modus:
  \begin{itemize}
    \item $read$ according to catalogue permissions
    \item $write$-$once$ according to catalogue permissions
    \item no directory manipulation allowed
   \end{itemize}
\end{itemize}

After evaluation of existing (Grid) protocols and I/O servers like in section
\ref{protocols} a conclusion was that none of these solutions can
satisfy approximately the above requirements.\\
Motivated by the given requirements a new architecture for the Grid
file system transport layer has been developed: the ${\bf
  C}ross{\bf l}ink-{\bf C}ache~Architecture$.

\subsection{Crosslink-Cache Architecture}
\begin{figure}
\includegraphics[width=65mm]{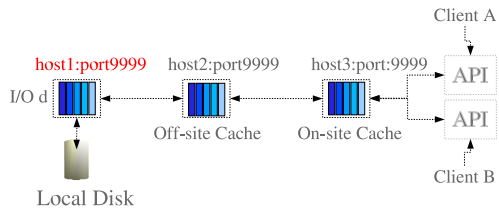}
\caption{Schematics of the Crosslink-Cache Architecture. I/O is routed
through on- and off-site caches.}
\label{CLC}
\end{figure}

Since Grid files are shared between a large community of users, it is
reasonable to move from a point-to-point connection scheme between
SEs and off-site applications to connections of a cache-gateway
type: ${\bf ClC}~Architecture$.\\
In this scheme file accesses are routed through distributed cache
site-servers to allow efficient usage of cache functionality (see
figure \ref{CLC}).
A Grid service will provide for each file access information about the
most efficient file access route. Moreover this service can route file
$write$ operations through caches to perform automatic file replication.\\
To allow asynchronous caching, each file has to be 
labeled with a globally unique identifier GUID \cite{GUID}. The {\it
  write-once}
strategy combined with GUID labeling guarantees then the identity of files
with the same GUID label in different caches.\\
\subsubsection{$aiod$ Cache-And-Forward Server}
\begin{figure}
\includegraphics[width=65mm]{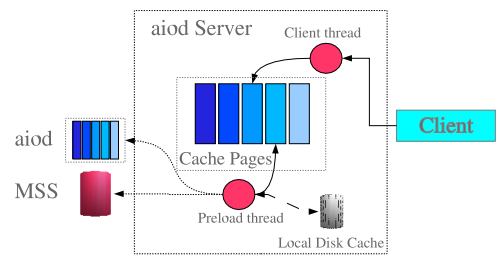}
\caption{Schematics of the $aiod$ Cache-and-Forward Server.}
\label{AIOD}
\end{figure}

At present a new I/O server $aiod$ is under development to allow the
cache-and-forward mechanism (see figure \ref{AIOD}). It consists of two daemons, one handling
the client $read/write$ requests and one pre-loading the clients data
from other $aiod$ servers or local storage systems. The transfers are
split into cache pages depending on the total file size and client
request (f.e. random access requires preferably small cache pages).\\
The authentication is done with certificates, which reside in the
client directory inside the
AliEn file catalogue as virtual DB files. Each cache page can be
encrypted using the SSL envelope mechanism \cite{SSL} for efficient performance .
For the transport between servers the QUANTA extended parallel TCP
class has been modified to allow transfer speed regulation
\cite{QUANTA}. A C++ client class and an $aioget$ and $aioput$
command has been developed for file access from applications and
shells. If a file is requested by a client, the client has to provide the
default routing information (cache-forward addresses like $host1@host2@host3$), Grid username, certificate file
locations, LFN, PFN, GUID, encryption type to the $aiod$ server. This
information is validated by the $aiod$ server before the access is
performed. All transfered cache pages are written with GUID, offset
and size information into the cache directory. In concurrent read
requests, data can be taken out of the cache directory instead of
downloading from
remote servers through the network.

\subsubsection* {Site I/O Gate Keeper}
\begin{figure}
\includegraphics[width=65mm]{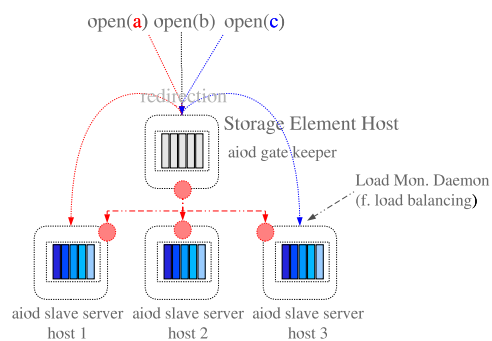}
\caption{The $aiod$ gate keeper redirects incoming I/O requests to
  slave servers.}
\label{Gatekeeper}
\end{figure}

If an $aiod$ server is set up as a I/O gate keeper, it can re-route a
connection to a slave server (see figure \ref{Gatekeeper}). The total I/O bandwidth
and status of all configured slave servers are reported through a
monitoring daemon to the gate keeper, which chooses the server with
the lowest load.\\

\subsubsection* {Site Cache Gate Keeper}
The same $aiod$ server can act as a cache gate keeper. If a client
asks for a special GUID, all slave cache servers are queried by the
gate keeper for this GUID. In case of a match the client is
redirected to the associated slave cache.\\

\subsubsection* {Site-Setup with Crosslink-Cache Architecture}
Ideally every site should be configured with cache gate keepers,
which present the access point for all off-site and on-site I/O
requests. These gate keepers redirect to slave caches. If files are
not cached the slave cache forwards from the I/O gate keeper, which
redirects to a slave I/O server with a low load. Data is then read
by a slave I/O server through a slave I/O cache server. Ideally the slave
cache itself acts also as a slave server and the cache gate keeper as
the I/O gate keeper.\\

\section {Operational areas and setup of the Global Grid File System in AliEn}
A combination of the concepts introduced in the
previous sections points out a possible scenario for the use of
Grid File Systems in multi-user environments. Interactive work and
applications use the Grid File System in a slightly different way\\
\subsection {Interactive Work}
The needs of interactive work are fully satisfied by the VFS
implementation. Users can read, edit and write files of the Grid
File System like in every conventional file system interactively. \\
\begin{figure}
\includegraphics[width=70mm]{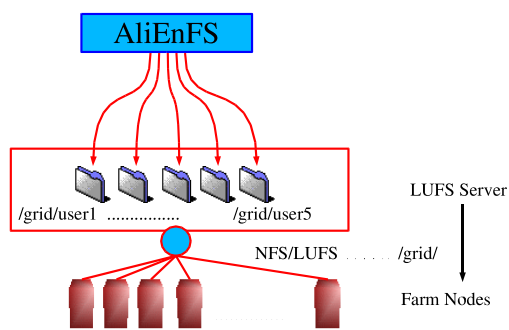}
\caption{Grid File System Export via NFS/LUFS modules from a LUFS server.}

\label{LUFSExport}
\end{figure}
A special setup can allow to cross-mount the Grid File System from a
LUFS server via NFS or LUFS modules to Linux nodes or via Samba to
Windows computers (see figure \ref{LUFSExport}). The LUFS server can act in this scenario moreover as a
user-shared file cache. The export of the Grid File System with user
specific permissions will require no additional kernel modules on client
nodes and simplifies the installation of large computing farms.\\
Each user has to setup once the
authentication for the LUFS server to allow auto-mounting of the Grid file
catalogue with specific user permissions. The security of the export from the LUFS
server to any file system client depends then on the used
protocol (Samba/NFS/etc.).\\
The unix-like command line interface allows all standard file system operations.
The syntax \\
{\hspace{1mm}}
 ``{\it $<$command$>$ $<$file1$>$@$<$SE1$>$
  $<$file2$>$@$<$SE2$>$}'' allows to implement all additional file moving, linking and replication commands
of a Grid File System.\\
All transfers from the Grid File System to any other local file system
or the other way around have to be executed synchronously. On the other hand
replication requests of large files (f.e. {\it cp $<$LFN$>$@$<$SE1$>$ $<$LFN$>$@$<$SE2$>$})
between storage elements can be executed asynchronously. These requests can be queued in a file transfer
queue. When the transfer is scheduled and executed, the new replica
location will be added to the LFN.\\

\subsection {Applications}
\begin{figure*}
\includegraphics[width=140mm]{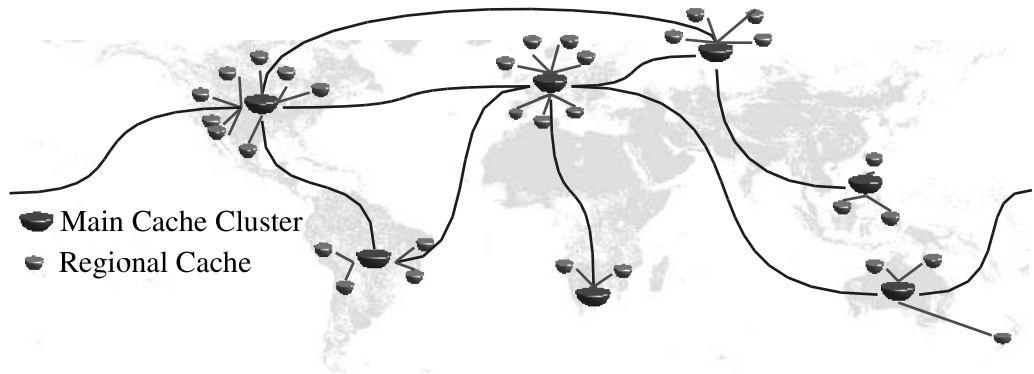}
\caption{Transfer network with main and regional cache
  clusters. Cross-continental data transfers are routed through main
  cache clusters.}
\label{WCache}
\end{figure*}
Applications use the AliEn API library for file access. The node setup
requires only the installation of the AliEn API libraries.\\
Access should be routed through regional and main cache servers to
minimise data movements. Regional or nationwide cache servers are
connected through high-bandwidth networks like shown in figure \ref{WCache}.\\

\subsection {File Replication}
File replication is done in an active and passive way. Files are
replicated by explicit replication requests using the file transfer queue.
On the other hand repetitive file access triggers an automatic file replication
into a nearby storage element. In this case replication needs no
additional long-distance transfers anymore, since accessed files are already cached in regional cache servers.

\section {Summary and Outlook}
The implementation of a Global Grid File System has importancy for
any Grid platform. The AliEn ambition is to provide a self-made server for the global transport layer and to use existing
protocols for local transfers. The user can use the file system
through a VFS and a {\it gridfs} module like a conventual file system,
while applications will make use of the AliEn API. AliEn will ideally
also support other Grid Storage solutions like the EDG module, to glue
most existing solutions in one. \\
The Crosslink-Cache architecture will allow cached file access and
implements the
idea of efficient resource sharing and resource control. \\
The future use in the ALICE experiment will assert or disprove the
conceptual design very soon.\\



\end{document}